\documentclass[preprint,aps,prd,showpacs,nofootinbib,superscriptaddress,
tightenlines,
floatfix]{revtex4-1}

\usepackage{slashed}
\usepackage{amsmath}
\usepackage{graphicx}
\usepackage{epstopdf}
\usepackage{epsfig}
\usepackage{amssymb}
\usepackage{bm}
\newcommand{\beq}{\begin{equation}}
\newcommand{\eeq}{\end{equation}}
\newcommand{\ov}{\overline}

\usepackage{color}

\begin{document}

\begin{flushright}
{\small  IPMU--15--0030} \\
\end{flushright}

\title{ Flavor violating $Z'$ from $SO(10)$ SUSY GUT in High-Scale SUSY}

\author{Junji Hisano}
\affiliation{Department of Physics, Nagoya University, Nagoya 464-8602, Japan}
\affiliation{Kavli IPMU (WPI), University of Tokyo, Kashiwa, Chiba 277-8583, Japan}

\author{Yu Muramatsu }
\affiliation{Department of Physics, Nagoya University, Nagoya 464-8602, Japan}
\author{Yuji Omura}
\affiliation{Department of Physics, Nagoya University, Nagoya 464-8602, Japan}
\author{Masato Yamanaka }
\affiliation{Department of Physics, Nagoya University, Nagoya 464-8602, Japan}


\begin{abstract} 
  We propose an $SO(10)$ supersymmetric grand unified theory (SUSY
  GUT), where the $SO(10)$ gauge symmetry breaks down to $SU(3)_c
  \times SU(2)_L \times U(1)_Y\times U(1)_{X}$ at the GUT scale and
  $U(1)_X$ is radiatively broken at the SUSY-braking scale.  In order
  to achieve the observed Higgs mass around $126$~GeV and also to satisfy
  constraints on flavor- and/or CP-violating processes, we assume that
  the SUSY-breaking scale is $O(100)$~TeV, so that the $U(1)_X$
  breaking scale is also $O(100)$~TeV.  One big issue in the SO(10)
  GUTs is how to realize realistic Yukawa couplings. In our model, not
  only ${\bf 16}$-dimensional but also ${\bf 10}$-dimensional matter
  fields are introduced to predict the observed fermion masses and
  mixings.  The Standard-Model quarks and leptons are linear
  combinations of the ${\bf 16}$- and ${\bf 10}$-dimensional fields so
  that the $U(1)_{X}$ gauge interaction may be flavor-violating.  We
  investigate the current constraints on the flavor-violating $Z'$
  interaction from the flavor physics and discuss prospects for future
  experiments.
\end{abstract}



\maketitle

\section{Introduction}
The Grand Unified Theories (GUTs) are longstanding hypotheses, and
continue to fascinate us because of the excellent explanation of
mysteries in the Standard Model (SM).  The GUTs unify not only the
gauge groups but also quarks and leptons, and reveal the origin of the
structure of the SM, such as the hypercharge assignment for the SM
particles.

The gauge groups in the SM are $ SU(3)_c \times SU(2)_L \times U(1)_Y
(\equiv G_{SM})$.  The minimal candidate for the unified gauge group
is $SU(5)$, which was originally proposed by Georgi and Glashow
\cite{GG}. There, quarks and leptons belong to ${\bf 10}$- and
$\ov{{\bf 5}}$-dimensional representations in $SU(5)$, and the SM
Higgs doublet is embedded into ${\bf 5}$, introducing additional
colored Higgs particle.  One big issue is the unification of the SM
gauge coupling constants, and it could be realized in the
supersymmetric (SUSY) extension.  It is well-known that the minimal
$SU(5)$ SUSY GUT realizes the gauge coupling unification around $2 \times
10^{16}$~GeV, if SUSY particle masses are around $1$~TeV
\cite{su5}.

Another candidate for the unified gauge group would be $SO(10)$. It is
non-minimal, but it would be an attractive extension because the
$SO(10)$ GUT explains the anomaly-free conditions in the
SM. Furthermore, all leptons and quarks, including the right-handed
neutrinos, in one generation may belong to one ${\bf 16}$-dimensional
representation in the minimal setup \cite{so10}. 

On the other hand, the GUTs face several problems, especially because
of the experimental constraints.  One stringent constraint is from
nonobservation of proton decay \cite{GG,dim5protondecay}.
While the GUT scale in the SUSY GUT may be high enough to suppress the proton
decay induced by the so-called $X$-boson exchange, the
dimension-five operator generated by the colored Higgs exchange is
severely constrained. Another stringent constraint is from the
observed fermion masses and mixings.  The $SU(5)$ GUT predicts a
common mass ratio of down-type quark and charged lepton in each
generation. Furthermore, in the $SO(10)$ GUT, the up-type, down-type
quarks, and charged lepton in each generation would have common mass ratios if
the all matter fields in one generation are embedded in one ${\bf
  16}$-dimensional representation.  The predictions obviously conflict
with the observation, and the modifications should be achieved by, for
instance, higher-dimensional operators \cite{FermionMass via HDO},
additional Higgs fields \cite{FermionMass via EH} and 
additional matter fields \cite{Barr:1981wv}.

In this letter, we propose an $SO(10)$ SUSY GUT model, where the
realistic fermion masses and mixings may be achieved by introducing
extra ${\bf 10}$-dimensional matter fields.  The SM quarks and leptons
come from ${\bf 10}$- and ${\bf 16}$-dimensional fields, and
especially, the right-handed down-type quarks and left-handed leptons in
the SM are given by the linear combinations of ${\bf 10}$- and ${\bf
  16}$-dimensional fields. We assume that $SO(10)$ gauge symmetry
breaks down to $G_{SM} \times U(1)_{X}$ around $10^{16}$~GeV according to
the nonzero vacuum expectation values (VEVs) of $SO(10)$ adjoint
fields. Thus, the low-energy effective theory is an $U(1)_{X}$
extension of the SUSY SM with extra matters.  The additional gauge
symmetry will survive up to the SUSY scale, but we could expect that
it is radiatively broken, as the electroweak (EW) symmetry breaking in
the minimal supersymmetry Standard Model (MSSM).

We assume that SUSY particles in the SUSY SM, except for gauginos,
reside around $100$~TeV, in order to realize the observed $126$~GeV
Higgs mass and also to satisfy constraints on flavor- and/or
CP-violating processes. This type of setup is called the high-scale SUSY
\cite{HSSUSY}.  In the high-scale SUSY, the gauge coupling unification
is rather improved when only the gaugino masses are around $1$~TeV
\cite{GCU in split SUSY}, and the dangerous dimension-five proton decay
is suppressed \cite{Hisano:2013exa}. On the other hand, since $\tan\beta$
(the ratio of the VEVs of the two Higgs doublets
in the SUSY SM) is close to one,  it is difficult to explain the
large hierarchy between top and bottom quarks when all the matter
fields are embedded into only ${\bf 16}$ representational
representations. In our model, the introduction of ${\bf 10}$-representational matter fields makes it possible to explain the large
hierarchy. In the high-scale SUSY, the UV theory of the SM need not be the MSSM.  
The $U(1)_{X}$ extension of the SUSY SM
with extra matters is an alternative model, motivated by the $SO(10)$ SUSY
GUTs.

The mass of the $Z'$ boson associated with the gauged $U(1)_{X}$ may
be $O(100)$~TeV so that it may be viable in the searches for flavor
violations.  The right-handed down-type quarks and left-handed leptons in
the SM are given by linear combinations of the parts of ${\bf 10}$- and
${\bf 16}$-dimensional fields. Thus, that generically leads
flavor-violating $Z'$ interaction and crucial promises against flavor
experiments.  We will see that tree-level Flavor Changing Neutral
Currents (FCNC) induced by the $Z'$ boson are generated and they
largely contribute to the flavor violation processes: for instance,
$\mu \to 3 e$, $\mu$-$e$ conversion in nuclei, and $K^0-\ov{K}^0$ and
$B_{d/s}^0-\ov{B}^0_{d/s}$ mixings.
 
This paper is organized as follows. We introduce our setup of the
$SO(10)$ SUSY GUT model in Sec. \ref{sec1}.  We see not only how to
break $SO(10)$, but also how to realize realistic fermion masses and
mixings.  The conventional seesaw mechanism, in which the Majorana
masses for the right-handed neutrinos are much higher than the EW
scale, could not work, since the $U(1)_{X}$ gauge symmetry forbids the
Majorana masses.  We show our solution according to the so-called
inverted hierarchy \cite{InverseSeesaw} in the Sec.~\ref{sec2-2}.  The
small parameters could be controlled with the global $U(1)_{PQ}$
symmetry there. In Sec.~\ref{sec2-3}, we discuss the tree-level FCNCs
corresponding to the realistic fermion masses and mixings.
Sec.~\ref{Sec:Flavor} is devoted to the flavor physics induced by the
$Z'$ interaction.  Sec.~\ref{sec4} is conclusion and discussion.

\section{Setup of $SO(10)$ SUSY GUT}
\label{sec1}
The $SO(10)$ gauge group has been considered to unify the three gauge
symmetry in the SM.  In the simple setup, the SM matter fields are
also unified into ${\bf 16}$-dimensional representation in the each generation, and
the number of Yukawa couplings for the fermions masses is less than in
the SM.  When the SM Higgs field belongs to ${\bf 10}$-dimensional field ${\bf 10}_H$, the
only Yukawa couplings are \beq W_{\rm min}=h_{ij} {\bf 16 }_i {\bf
  16}_j {\bf 10}_H\,  \eeq where $i,j=1, \,2,\,3$ are defined. This assumption is too strict
to explain the observed fermion masses and mixings, even if we include
radiative corrections.  The observed mass hierarchies are different in
the up-type and down-type quarks, and the CKM mixing will be vanishing 
without other Yukawa couplings.

Now, we introduce a ${\bf 10}$-dimensional matter field in the each
generation in addition to ${\bf 16}$-dimensional matter fields.  Three
$SO(10)$-singlet matter fields $S_i$ are also introduced to
achieve the realistic masses of neutrinos.  The matter fields ${\bf
  10}_i$ and ${\bf 16}_i$ are decomposed as the ones in Table~\ref{table2}.  For
convenience, the assignment of $SU(5) \times U(1)_X$ is also shown in
Table~\ref{table2}.

\begin{table}[t]
\begin{center}
\begin{tabular}{|c|c|c|c|c|c|c|}
\hline
 & $Q_L$ & $U_R^c$ & $E_R^c$ & $\hat L_L$ & $\hat D_R^c$ & $N_R^c$ \\ \hline
$SO(10)$ & \multicolumn{6}{|c|}{${\bf 16}$} \\ \hline
$SU(5) \times U(1)_X$ & \multicolumn{3}{|c|}{$({\bf 10},-1)$} & \multicolumn{2}{|c|}{$({\bf \bar 5},3)$} & 
$({\bf 1},-5)$ \\ \hline
$G_{\rm SM}$ & $({\bf 3},{\bf 2},\frac{1}{6})$ & $({\bf \bar 3},{\bf 1},-\frac{2}{3})$ & 
$({\bf 1},{\bf 1},1)$ & $({\bf 1},{\bf 2},-\frac{1}{2})$ & $({\bf \bar 3},{\bf 1},\frac{1}{3})$ & 
$({\bf 1},{\bf 1},0)$ \\ \hline \hline
 & ${L'}_L$ & ${D'}_R^c$ &  $\overline{{L'}_L}$ & $\overline{{D'}_R^c}$ \\ \cline{1-5}
$SO(10)$ & \multicolumn{4}{|c|}{${\bf 10}$} \\ \cline{1-5}
$SU(5)\times U(1)_X$ & \multicolumn{2}{|c|}{$({\bf \bar 5},-2)$} & \multicolumn{2}{|c|}{$({\bf 5},2)$}
\\ \cline{1-5}
$G_{\rm SM}$ & $({\bf 1},{\bf 2},-\frac{1}{2})$ & $({\bf \bar 3},{\bf 1},\frac{1}{3})$ & 
$({\bf 1},{\bf 2},\frac{1}{2})$ & $({\bf 3},{\bf 1},-\frac{1}{3})$ \\ \cline{1-5} \cline{1-5}
\end{tabular} 
\caption{
\label{table2}%
{
Charge assignment for matter fields. 
Charge assignment for $G_{\rm SM}$ is denoted as $(SU(3)_c$, $SU(2)_L$, $U(1)_Y$).
$U(1)_X$ gauge coupling constant is normalized as $g_X=g/\sqrt{40}$ 
at GUT scale, where $g$ is $SO(10)$ gauge coupling constant. 
}
}
\end{center}
\end{table}

Let us show the superpotential relevant to the Yukawa couplings for
the matter fields in our model;
\begin{eqnarray}
W_{Y}&=&h_{ij} {\bf 16 }_i  {\bf 16}_j {\bf 10}_H +  f_{ij} {\bf 16}_i  \overline{\bf{16}}_H S_j +
g_{ij}  {\bf 10}_i{\bf 16}_j {\bf 16}_H \nonumber \\
&&+\mu_{BL} {\bf 16}_H  \overline{\bf{16}}_H +\mu_H {\bf 10}_H {\bf 10}_H   + 
\mu_{10\,ij} {\bf 10}_i  {\bf 10}_j  + \mu_{S\,ij} S_i S_j. \label{superpotential2}
\end{eqnarray}
Here, the ${\bf 16}$ and $\ov{{\bf 16}}$-dimensional Higgs fields
${\bf 16}_H$ and $\ov{{\bf 16}}_H$ are introduced to break the
$U(1)_X$ gauge symmetry in $SO(10)$. We assume that the mass
parameters $\mu_{BL}$, $\mu_{10}$ and $\mu_H$ are around SUSY scale
$(m_{SUSY})$ and $\mu_S$ is much smaller to realize the tiny neutrino
masses.  It may be important to pursue the origin of the mass scales.
In Sec. \ref{sec2-2}, we show that the global $U(1)_{PQ}$ symmetry may
control their mass scales.

We assume that two $SO(10)$ adjoint Higgs fields, ${\bf 45}_H$ and
${\bf 45}'_H$, develop nonzero VEVs so that the $SO(10)$ gauge symmetry
breaks down to $G_{SM} \times U(1)_{X}$ at the GUT scale \cite{SO(10)
  breaking chain}. The low-energy effective theory is the $U(1)_{X}$
extension of the SUSY SM with ${\bf 10}$- and ${\bf
  16}$-dimensional matter fields.  The $G_{SM}$-singlet fields charged
under $U(1)_{X}$, $\Phi$ and $\ov{\Phi}$, which are originated from
${\bf 16}_H$ and $\ov{{\bf 16}}_H$, should be included there.  $\Phi$
and $\ov{\Phi}$ would develop the nonzero VEVs as $\langle \Phi
\rangle=v_\Phi$ and $\langle \ov{\Phi} \rangle =\ov{v_\Phi}$ around
$m_{SUSY}$, and the $U(1)_{X}$ symmetry is spontaneously broken.  For
simplicity, we assume that the other fields in ${\bf 16}_H$ and $\ov{{\bf
    16}}_H$ have masses at the GUT scale.  If they stay at the low
energy spectrum, the gauge coupling constants at the GUT scale is
not perturbative.

The superpotential in the $U(1)_{X}$
extension of the SUSY SM is given as follows,
\begin{eqnarray}
W^{eff}_Y &=&h_{u\,ij} Q_{L\,i} U_{R\,j}^{c} H_u + 
(h_{u\,ij} +\epsilon_{d\,ij}) Q_{L\,i} \Hat D_{R\,j}^{c} H_d + 
(h_{u\,ij} +\epsilon_{e\,ij}) \Hat L_{L\,i} E_{R\,j}^{c} H_d \nonumber  \\
&&+ g_{ij} \Phi  (\ov{{D'}_R^c}_i \Hat D_{R\,j}^{c}+\ov{{L'}_L}_i \Hat L_{L\,j})+ 
\mu_{10\,ij}  (\ov{{D'}_R^c}_i  {D'}_{R\,j}^{c}+\ov{{L'}_L}_i  {L'}_{L\,j}) 
\nonumber \\
&& +h_{ij} \Hat L_{L\,i} N^{c}_{R\,j} H_u +
f_{ij} \ov{\Phi}N^{c}_{R\,i} S_j + 
\mu_{S \,ij} S_i S_j +\mu_{BL} \ov{\Phi}\Phi +\mu_H H_uH_d. \label{superpotential3} 
\end{eqnarray}
The effective Yukawa couplings will be deviated from the ones in
Eq.~(\ref{superpotential2}), because of the higher-order terms
involving ${\bf 45}_H$ and ${\bf 45}'_H$.\footnote{
In general, the other parameters such as $\mu_S$ and $\mu_{10}$ would be
effectively modified by the higher-dimensional operators as well.  We
disregard these extra corrections to the parameters because they are
not essential in this discussion.
}  $h_u$ is Yukawa coupling
for up-type quark including effect of higher-dimensional operators.
$\epsilon_d$ and $\epsilon_e$ describe the size of higher-dimensional
operators for the down-type quarks  and charged leptons, which
suppressed by $\langle {\bf 45}_H \rangle/\Lambda $ and
$\langle {\bf 45}'_H \rangle/\Lambda $.

After the $U(1)_X$ symmetry breaking, the chiral superfields 
$\Hat D^{c}_{R\,i}$ and ${D'}^{c}_{R\,i}$ $(\Hat L_{L\,i}$ and ${L'}_{L\,i})$
mix each other, and we find the massless modes which correspond to the SM right-handed down-type quarks and left-handed leptons.  $g_{ij} v_\Phi$ and $\mu_{10\,ij}$ 
give the mass mixing between $\Hat D^{c}_{R\,i}$ and ${D'}^{c}_{R\,i}$ ($\Hat L_{L\,i}$ and ${L'}_{L\,i}$).
Eventually, the relevant Yukawa couplings for quarks and charged leptons are described as 
\beq
W^{SSM}_Y=h_{u\,ij} Q_{L\,i} U_{R\,j}^{c} H_u + 
Y_{d\,ij}  Q_{L\,i}  D_{R\,j}^{c} H_d + 
Y_{e\,ij}   L_{L\,i} E_{R\,j}^{c} H_d  + 
\widetilde \mu_{ij} ({\ov{D_{R\,h}^c}_i  D^{c}_{R\,h\,j} + \ov{L_{L\,h}}_i  L_{L\,h\,j}}).
 \label{superpotentialMSSM}
\eeq
$D^{c}_{R\,i}$, $D^{c}_{R\,h\,i}$, $L_{L\,i}$ and $L_{L\,h\,i}$ are the chiral superfields of right-handed down-type quarks and left-handed leptons in the mass bases defined by
\begin{equation}\label{massbase}
\begin{pmatrix} \Hat \psi \\ \psi'  \end{pmatrix} =
U_{\psi}   \begin{pmatrix} \psi \\ \psi_h \end{pmatrix}=  \left(
\begin{array}{cc}
\Hat U_{\psi} & \Hat U_{\psi\,h}  \\
\Hat U'_{\psi} & \Hat U'_{\psi\,h}
\end{array}
\right)  \begin{pmatrix} \psi \\ \psi_h \end{pmatrix},
\end{equation}
where $\psi$ denotes $D_R^c$ and $L_L$.
$\psi$ and $\psi_h$ are massless modes which correspond to the SM matters and the superheavy modes with masses
$O(m_{SUSY})$, respectively.
$U_{\psi}$ is the $6 \times 6$ unitary matrix, and $\Hat U_{\psi}$, $\Hat U_{\psi\,h}$, $\Hat U'_{\psi}$ and 
$\Hat U'_{\psi\,h}$ satisfy not only the unitarity condition 
for $U_{\psi}$ but also the following relation,
\begin{eqnarray}
0&=&g_{ik} v_\Phi( \Hat U_{\psi})_{kj} + \mu_{10\,ik} ( \Hat U'_{\psi})_{kj}, \\
\widetilde \mu_{ij}&=&g_{ik}v_\Phi ( \Hat U_{\psi\,h})_{kj} + \mu_{10\,ik} ( \Hat U'_{\psi\,h})_{kj}.
\end{eqnarray}
Using the couplings in Eq.~(\ref{superpotential3}), the Yukawa coupling constants for the SM down-type quarks and charged leptons in Eq.~(\ref{superpotentialMSSM}) are described as
\beq 
\label{downYukawa}
(Y_d)_{ij}=(h_{u\,ik} + \epsilon_{d\,ik}) (\Hat U_{D^c_R})_{kj},~
(Y_e)_{ij}=(\Hat U_{L_L}^T)_{ik}(h_{u\,kj} + \epsilon_{e\,kj}) .
\eeq
In general, the up-type quark Yukawa coupling constants $h_{u\,ij}$ is given by
\beq
h_{u\,ij}= \frac{m_{u\,i}}{v \sin \beta}\delta_{ij}.
\eeq
$v \sin \beta$ ($v \cos \beta$) is the VEV of the neutral component of $H_u$ $(H_d)$ and $m_{u\,i}$ are the up-type quark masses.
We define the diagonalizing matrices $V_{CKM}$ and $V_{e_R}$ for $(Y_d)_{ij}$ and $(Y_e)_{ij}$ as below:
\begin{equation}
(Y_{d})_{ij}= \frac{1}{v \cos \beta}(V^*_{CKM})_{ij} 
 m_{d\,j} ,~ (Y^T_{e})_{ij}= \frac{1}{v \cos \beta}(V^*_{e_R})_{ij} 
 m_{e\,j},
\end{equation}
where $m_{d\,i}$ and $m_{e\,i}$ are the down-type quark and the
charged lepton masses.  Note that we take the flavor basis that the
right-handed down-type quarks and left-handed charged leptons are in
the mass eigenstates. Then $V_{CKM}$ is the CKM matrix
and $V_{e_R}$ satisfies $V_{e_R}=V_{CKM}$ in the $SU(5)$ limit.  

The size of higher-dimensional terms is depicted by $\epsilon_d$ and $\epsilon_e$ and expected to be small, compared to the third generation,
$h_{u\,33} =m_t/(v \sin \beta)$. 
According to Eq.~(\ref{downYukawa}), $(\Hat U_{\psi})_{ij}$ could be described by the observables as,
\begin{eqnarray} \label{ud}
 \left( m_{u\,i} \delta_{ik}+ \epsilon_{d\,ik} v \sin \beta \right) (\Hat U_{D_R^c})_{kj}&=&
 \tan \beta (V^*_{CKM})_{ij} 
 m_{d\,j}, \nonumber \\
 \left( m_{u\,i} \delta_{ik}+ \epsilon^{T}_{e\,ki} v \sin \beta \right) (\Hat U_{L_L})_{kj}&=&
 \tan \beta (V^*_{e_R})_{ij} 
 m_{e\,j}. \label{ud-2}
\end{eqnarray}

If $\epsilon_{d\,11}  v \sin \beta$ is sufficiently smaller than $m_u$, 
the $(1, \, j)$ elements of $\Hat U_{D_R^c}$ are too large to satisfy the unitary condition for $U_{\psi}$.
In order to achieve the consistency, the extra term $\epsilon_{d\,11} v \sin \beta$ should be larger than 
$O(\tan \beta  (V_{CKM})_{13}m_b)$.

\subsection{Neutrino Mass}
\label{sec2-2}
Let us briefly mention the neutrino sector in our model.
$W^{eff}_Y$ in Eq.~(\ref{superpotential3}) includes neutral particles after the EW symmetry breaking.
They reside in the neutral components of $SU(2)_L$ doublets $\{ \Hat L_{L\,i}, \,{L'}_{L\,i}, \, 
\ov{{L'}_L}_i \}$ and the singlets \{$N_{R\,i}$, $S_i$\}.
Let us decompose $\Hat L_{L\,i}$, ${L'}_{L\,i}$ and $\ov{{L'}_L}_i$ as the charged and neutral ones:
$\Hat L^T_{L\,i}=( \Hat \nu_{L\,i}, \, \Hat E_{L\,i})$, ${L'}^T_{L\,i}=({\nu'}_{L\,i},\, {E'}_{L\,i})$ and 
$\ov{{L'}_L}^T_i=(\ov{{\nu'}_L}_i, \, \ov{{E'}_L}_i)$. 
The mass matrix for the neutral particles in the basis of
$( \Hat \nu_{L\,i}, \, N^{c}_{R\,i},  \, {\nu'}_{L\,i}, \, \ov{{\nu'}_L}_i, \, S_i)$ is
\begin{equation}
M_{\nu} =\begin{pmatrix} 0 & h_{ij} v \sin \beta  & 0 &   g_{ij} v_\Phi & 0\\
 h_{ij} v \sin \beta &   0 &   0 & 0 & f_{ij} \ov{v_{\Phi}} \\
0 &   0 & 0   &  \mu_{10 \,ij} &   0   \\
 g_{ij} v_\Phi  &  0 &  \mu_{10 \,ij}  &  0 &  0  \\
  0 & f_{ij} \ov{v_\Phi} &  0 & 0  & \mu_{S\,ij}  \end{pmatrix}.
\end{equation}
When we admit the large hierarchy between $\mu_S$ and
the other elements, the neutrino mass matrix $(m_\nu)$ is given by
\beq
(m_{\nu})_{ij}=  (hf^{-1} \mu_S f^{-1}h)_{ij}  \left( \frac{v \sin \beta}{ \ov{v_\Phi}} \right)^2,
\eeq
following Ref. \cite{InverseSeesaw}.
For instance, $\ov{v_\Phi}=O(100)$~TeV and $v \sin \beta=O(100)$~GeV lead  $O(1)$-eV neutrino masses, if $\mu_S$ is $O(1)$ MeV and $h$ and $f$ are $O(1)$.
The masses of the other neutral elements are $O(m_{SUSY})$, and
the phenomenology has been well investigated in Ref. \cite{InverseSeesaw}.

\begin{table}[th]
\begin{center}
\begin{tabular}{c|cccc|| c c || c c}
     & ~${\bf 16}_i $~ & ~${\bf 10}_H$~ &~${\bf 16}_H $~ &~$\ov{{\bf 16}}_H $~ & ~$10_i$ ~ &~ $S_i$~ &~$ P$~ &  ~$T$      \\ \hline  
$SO(10)$  & \bf{16} & \bf{10} & \bf{16} & $\ov{\bf{16}}$  &  \bf{10}& \bf{1} & \bf{1}  & \bf{1}  \\ \hline
$U(1)_{PQ}$  & 1      & -2          & -1/3   & 5/3    &  -2/3        & -8/3        & -2/3  & 2  \\ 
\end{tabular}
\caption{
\label{table1}%
{
Charge assignment of global $U(1)_{PQ}$ symmetry. 
}
}
\end{center}
\end{table}

One may wonder why $\mu_S$ is so tiny and $\mu_{10,BL,H}$ are $O(m_{SUSY})$. We show one mechanism to explain the large mass hierarchy.
In order to induce the dimensional parameters in Eq.~(\ref{superpotential2}) effectively,
let us assign the global $U(1)_{PQ}$ symmetry to the matter and Higgs fields as in Table \ref{table1}.
The global $U(1)_{PQ}$ symmetry, under which the SM fields are charged anomalously, has been proposed motivated by the strong CP problem \cite{PQ}. 
We introduce $SO(10)$-singlet fields, $P$ and $T$, whose $U(1)_{PQ}$ charges are fixed to allow the $c_{PQ} P^3T$ term in the superpotential. Assuming canonical K\"aller potential and their soft SUSY breaking terms, the scale potential for $P$ and $T$ is derived from the superpotential as
\beq
V_{PQ}=\left | \frac{c_{PQ}}{\Lambda} P^3 \right|^2 +  \left |\frac{c_{PQ}}{\Lambda} P^2 T \right|^2 +m^2_{P} |P|^2  +m^2_{T} |T|^2.
\eeq
$m^2_{P} $ and $m^2_{T} $ are the soft SUSY breaking masses, and they could be estimated as $m^2_{SUSY}$.
The mass squared would be driven to the negative value due to the radiative corrections, so that the negative mass squared leads the nonzero VEVs of $P$ and $T$,
\beq
\langle T \rangle \sim \langle P \rangle \sim \sqrt{\Lambda |m_{SUSY}|},
\eeq
and breaks $U(1)_{PQ}$ spontaneously. This leads a light scalar, so-called axion, corresponding to the
Nambu-Goldstone boson.
As discussed in Ref. \cite{axion}, it is favorable that the $U(1)_{PQ}$ symmetry breaking scale is around $10^{12}$~GeV, to explain the correct relic density of dark matter. That is, $\Lambda$ should be almost the Planck scale ($O(10^{19})$~GeV), when $m_{SUSY}$ is $O(100)$~TeV, for instance.

On the other hand, the $U(1)_{PQ}$ charge assignment for the other chiral superfields forbids 
dimensional parameters like $\mu_S$ and $\mu_{10,BL,H}$.
Using higher dimensional parameters, $\mu_S$ and $\mu_{10,BL,H}$ are effectively 
generated after $U(1)_{PQ}$ breaking:
\begin{equation}
\mu_{10} =  \frac{\langle P \rangle \langle T \rangle }{\Lambda},~ \mu_{BL} =   \frac{\langle P \rangle^2  }{\Lambda},~ \mu_{H} =   \frac{\langle T \rangle^2  }{\Lambda}, ~
\mu_S=  \frac{ \langle P \rangle \langle T \rangle^3   }{\Lambda^3},  
\end{equation}
ignoring the dimensionless couplings in front of the higher-order couplings.
The above estimation tells that  $\mu_{10,BL,H}= O(m_{SUSY})$ and $\mu_S = m_{SUSY} \times O(m_{SUSY}/\Lambda)$.
If $m_{SUSY} \ll \Lambda$ is satisfied, very small $\mu_S$, compared to $m_{SUSY}$, is predicted, 
and could realize the observed light neutrino masses, as we discussed above.

\subsection{Flavor Violating Gauge Interaction}
\label{sec2-3}

As we see above, the SM right-handed down-type quarks and left-handed leptons are
given by the linear combinations of quarks and leptons in ${\bf 10}$-
and ${\bf 16}$-dimensional matter fields, respectively.  Since the
fields in ${\bf 10}$ and ${\bf 16}$ representations carry different $U(1)_X$ charges,
the SM fields may have flavor-dependent $U(1)_X$ interaction.  

Let us see it more explicitly. The $U(1)_X$ gauge interactions 
of right-handed down-type quarks and left-handed leptons are
described in the interaction basis as
\begin{equation}
{\cal L}_g= -i g_X (3 \overline{\Hat \varphi}_i\slashed{Z}'  \Hat \varphi_i
-2 \overline{\varphi'}_i\slashed{Z}'  {\varphi'}_i),
\end{equation}
where the factors $3$ and $-2$ are $U(1)_X$ charges for the fermionic components
$\hat \varphi_i$ and $\varphi'_i$ of the chiral superfields $\Hat{\psi}_i$ and $\psi'_i$.
$Z'$ is the $U(1)_X$ gauge boson and $g_X$ is defined as $g_X=g/\sqrt{40}$ at GUT scale,
where $g$ is the $SO(10)$ gauge coupling constant.
We have obtained the mass eigenstates for the fermions in Eqs. (\ref{massbase}) and 
(\ref{downYukawa}).
Using the unitary matrix $U_{\psi}$, we define the flavor-violating couplings
${A}^{\varphi}_{ij}$ for the SM fermions as
\begin{equation} \label{FCNC}
{\cal L}_g =  -ig_{X} \ov{\varphi}_i 
 \left( 5(\hat U^\dagger_\psi \hat U_\psi)_{ij}
-2 \delta_{ij} \right) \slashed{Z}' \varphi_j 
\equiv -ig_{X}{A}^{\varphi}_{ij} \ov{\varphi}_i \slashed{Z}' \varphi_j,
\end{equation}
where $\varphi$ is the fermion component of the chiral superfield $\psi$ in the mass base and denotes right-handed down-type quark ($d_{R}^c$) and left-handed lepton ($l_{L}$).

Here we discuss the size of flavor violating couplings
$A_{ij}^{\varphi}$.  According to Eq.~(\ref{ud}), $(\hat U_{D^c_R})_{ij} $ 
and $(\hat U_{L_L})_{ij} $ 
are depicted by the observables in the SM.  The flavor violating
couplings $A^{\varphi}_{ij}$ depend on the parameters, $\epsilon_d$ and
$\epsilon_e$.  They are required to satisfy the unitary condition for
$U_{\psi}$, as discussed in Eqs. (\ref{ud}).  In
other words, they should be sizable in some elements, compared to
$h^u_{ij}=m_{u \, i}/v \cos \delta_{ij}$, in order to break the GUT
relation and to realize realistic mass matrices.
Assuming $\epsilon_{d \, ij} =\epsilon_{i} \delta_{ij}$, at least $\epsilon_1 \gtrsim O( 10^{-5})$
is required to compensate for the small up quark mass.

Let us show one example to demonstrate the size of the flavor violating coupling $A^{d^c_R}_{ij}$.
Assuming $\epsilon_1 \gtrsim O( 10^{-5})$ and $\epsilon_2=\epsilon_3=0$, $A^{d^c_R}_{ij}$ is approximately estimated as
\beq
A^{d^c_R}_{ij} \approx  \frac{5 \tan ^2 \beta m_{d \, i} m_{d \, j}}{\left| \epsilon_1 v \sin \beta \right|^2}  (V_{CKM})_{1i}   (V^*_{CKM})_{1j} -2 \delta_{ij}. 
\eeq
Setting the extra parameter to $\epsilon_1 = 5 \times 10^{-4}$, $A^{d^c_R}_{ij}$ is 
estimated as
\begin{equation}\label{eq:Ad}
\left(  A^{d^c_R}_{ij} \right ) \approx \left(
\begin{array}{ccc}
-1.9  & 0.6 & 0.3  \\
0.6 & 1.6   & 2.2    \\
0.3 & 2.2   & -0.3 
\end{array}
\right).
\end{equation}
We find that all elements of the flavor violating couplings are $O(1)$, so that we need 
careful analyses of their contributions to flavor physics, even if the $Z'$ boson is quite heavy.

Note that the alignment of $ A^{l_L}_{ij}$ differs from the one of $ A^{d^c_R}_{ij}$,
because of the different mass spectrum between charged leptons and down-type quarks.
In any case, however, the size of $ A^{l_L}_{ij}$ would be also $O(1)$, because of the small electron mass.
The detail analysis on the relation between the FCNCs and the realistic mass spectrum will be given 
in Ref. \cite{future}. In Sec. \ref{Sec:Flavor}, we introduce the flavor constraints relevant
to our model and scan the current experimental bounds and future prospects in
flavor physics.


\subsection{Gauge Coupling Unification}
Before phenomenology, let us briefly comment on the gauge coupling
unification and the predicted $Z'$ coupling ($g_{X}$).  As well-known,
the MSSM miraculously achieves the unification of the three SM gauge
couplings, if at least gaugino masses are close to the EW scale.  We
assume the SUSY mass spectrum, where gauginos reside around the
TeV-scale and the other SUSY particle masses are around $100$~TeV.
It is shown in Ref.~\cite{GCU in split SUSY} that the unification of the 
gauge coupling constants is improved compared with the MSSM with the 
SUSY particle masses $O(1)$~TeV.

Once we determine the $SO(10)$ gauge coupling at the GUT scale according to the gauge coupling unification, 
the $U(1)_X$ gauge coupling $g_{X}(\mu)$ is derived with the renormalization group equation at the one-loop level as 
\beq
4 \pi \alpha^{-1}_{X}(\mu)=4 \pi \alpha^{-1}_{G} \times 40+b_{X}
\ln \left ( \frac{\Lambda_G^2}{\mu^2} \right ),
\eeq
where $\alpha_X=g^2_X/(4 \pi)$ and $\alpha_G=g^2(\Lambda_G)/(4 \pi)$ are defined and
$\Lambda_G$ is the unification scale.
$b_{X}$ is fixed by the number of $U(1)_X$-charged particles from $\mu$ to $\Lambda_G$.
In our scenario, right-handed neutrinos,
additional three ${\bf 10}$s of $SO(10)$, and the $U(1)_X$ breaking Higgs fields  as well as MSSM particles contribute to $b_X$ between
$m_{SUSY} $ and $\Lambda_G$, so that they lead $b_X=426$.
At the scale $\mu=100$~TeV, $g_{X}$ is estimated as 
\begin{equation}
g_{X}(100\,{\rm TeV})=0.073,
\label{Eq:gX}
\end{equation}
where the GUT scale and the gauge coupling with $m_{SUSY}=100$ TeV are given by
\begin{equation}
\Lambda_G=8.7 \times 10^{15}\, {\rm GeV},\, \alpha_{G}=0.062.
\end{equation}

Note that the introduction of additional matter fields increases the gauge coupling constant at the GUT scale $\alpha_{G}$. Furthermore, heavier gaugino masses than the EW scale decrease the GUT scale $\Lambda_G$. 
This means that the proton decay rate may be enhanced in our model \cite{protondecay,GCU in split SUSY}, and could be tested at the future proton decay searches.

\section{Flavor physics}  \label{Sec:Flavor}  

As discussed in the subsection \ref{sec2-3}, the tree-level FCNCs involving the $Z'$ boson 
may be  promised in our model. The flavor changing couplings denoted by $A^{\varphi}_{ij}$
could be $O(1)$ in the all elements, as we see in Eq.~(\ref{eq:Ad}).
Here, we sketch the relevant constraints on the flavor-violating $Z'$ interactions and give prospects for future experiments.

In our model, the SUSY SM Higgs doublets are  charged under $U(1)_X$, 
so that their nonzero VEVs contribute to the $Z'$ mass ($m_{Z'}$) as well as the SM gauge bosons.
The $U(1)_X$ charges of Higgs doublets are $\pm 2$ respectively, and then 
the mass mixing between $Z$ and $Z'$ is  generated by the VEVs as well.
The mixing angle between $Z$ and $Z'$ is approximately estimated as
\begin{equation}
\begin{split}
   \sin \theta \simeq 4 \frac{g_{X}}{g_{Z}} \frac{m_{Z}^2}{m_{Z'}^2},
\end{split}      
\end{equation}
where $g_Z $ is the gauge coupling of $Z$ boson and $m_Z$ is the $Z$ boson mass.  $\sin \theta$ is
about $ 3.4 \times 10^{-7}$ when $Z'$ mass and coupling are fixed at $m_{Z'}=100$~TeV and $g_X=0.073$.  Since the mixing is quite small
as long as the $Z'$ mass is $O(100)$~TeV, we treat with $Z$ and $Z'$ as the
fields in the mass basis and discuss the mixing effect up to $O(
\theta^2)$.

The gauge interactions of $Z$ and $Z'$ and SM fermions are given by
\begin{equation}
\begin{split}
   \mathcal{L} &= 
   - i\left(  g_Z \cos\theta J_\text{SM}^\mu 
   + g_{X} \sin\theta J_\text{GUT}^\mu  
   \right) Z_\mu 
   - i\left(  g_{X} \cos\theta J_\text{GUT}^\mu 
   - g_Z \sin\theta J_\text{SM}^\mu  \right) Z'_\mu, 
   \label{Eq:Lag_mixing}
\end{split}      
\end{equation}
where $J_\text{SM}^\mu$ is the SM weak neutral current, and $J_\text{GUT}^\mu$ is defined by  
\begin{equation}
\begin{split}
   J_\text{GUT}^\mu 
   &= 
    A^{l_L}_{ij} \overline{l_L}_{i} \gamma^\mu l_{L\,j} 
   - A^{d_R^c}_{ij} \overline{d_R}_{i} \gamma^\mu d_{R\,j} 
   + \overline{e_R}_{i} \gamma^\mu e_{R\,i} 
   - \overline{q_L}_{i} \gamma^\mu q_{L\,i} 
   +  \overline{u_R}_{i} \gamma^\mu u_{R\,i}. 
   \label{Eq:J_GUT}
\end{split}      
\end{equation}
The fermions in $ J_\text{GUT}^\mu$ describe the fermionic components of the MSSM chiral superfields in the mass base denoted by the capital letters. 
The neutral current $J_\text{GUT}^\mu$ may significantly contribute to
flavor violating processes: $B_{d/s}^0$-$\overline{B}_{d/s}^0$ and 
$K^0$-$\overline{K}^0$ mixings, flavor-violating decays, and $\mu$-$e$
conversion in nuclei.  Below, we summarize the constraints relevant to
the $Z'$ interaction, and discuss the predictions in flavor physics.
Note that we ignore contribution from SUSY flavor violating processes,
because the sfermion masses are $O(100)$~TeV.

\subsection{Flavor Violating Decays of Leptons}  \label{Sec:FVD}  

First, let us discuss the contributions to flavor violating decays of
leptons.  There are two types of flavor violating decays in the
presence of $Z'$ FCNCs: one is three-body flavor violating decays $l_j
\to l_i l_k l_k $ and the other is radiative flavor violating decays
$l_j \to l_i \gamma$.  With the $Z'$ FCNCs, the three-body flavor
violating decays occur at the tree level, while the radiative flavor
violating decays occur at the loop level. The radiative flavor violating
decays have smaller rates by $O(10^{-3})$ than the tree-level decays.  
If flavor violating interactions stem from
both left- and right-handed lepton (quark) sector, there might be a
strong enhancement in radiative flavor violating decays via a
chirality flip on an internal heavy fermion~\cite{Murakami:2001cs}.
In our model, however, there exists no such an enhancement because
only left-handed lepton (right-handed quark) have the flavor violating
interactions.  Hence we focus on the three-body flavor violating
decays.

Let us discuss the $\mu \to 3e$ process.  The current upper bound on the branching
ratio of $\mu \to 3e$ is $1.0 \times 10^{-12}$~\cite{Bellgardt:1987du}
and future experimental limit is expected to be $1.0 \times
10^{-16}$~\cite{Blondel:2013ia}.  In this model the branching ratio of
$\mu \to 3e$ is evaluated as follows,
\begin{equation}
\begin{split}
   &\text{BR}(\mu \to 3e) 
   =    1.1 \times 10^{-15} \left( \frac{g_X}{0.073}\right)^4 \left( \frac{100\,\text{TeV}}{m_{Z'}} \right)^4 
   \left | A^{l_L}_{12} \right |^2
   \left\{  1 + 0.63 \left| 1-0.93 A^{l_L}_{11} \right|^2 \right\}. 
\end{split}      
\end{equation}
This is below the current experimental bound as long as $m_{Z'}$ is
$O(100)$~TeV.  It is also important to emphasize that $\text{BR}(\mu
\to 3e)$ in our scenario has an additive structure in last bracket,
and our prediction may yield to the stringent bound.  If we assume
$m_{Z'} = 100\, \text{TeV}$ and $A^{l_L}_{11}=-2$, the Mu3e experiment
will cover $\left| A^{l_L}_{12} \right| \lesssim 0.1$.

We also evaluate the branching ratios of other lepton flavor
violating decays, and we find that they are also much below the current
experimental upper bounds.

\subsection{$\mu$-$e$ Conversion in Nuclei} \label{Sec:mue_conv} 

The flavor violating coupling $A^{l_L}_{12}$ also gives rise to the
$\mu$-$e$ conversion process.  The SINDRUM-I\hspace{-1pt}I experiment,
which searched for the $\mu$-$e$ conversion signal with the Au target,
gave the upper limit on the branching ratio: $\text{BR}(\mu^-
\text{Au} \to e^- \text{Au}) <7 \times 10^{-13}$~\cite{Bertl:2006up}.
The DeeMe~\cite{Natori:2014yba} and the COMET-I~\cite{Kuno:2013mha}
will launch soon and they aim to reach to $O(10^{-15})$ for the
branching ratio with different targets.  Furthermore, COMET-I\hspace{-1pt}I and  Mu2e
\cite{Bartoszek:2014mya} are planed to improve the sensitivity up to
$O(10^{-17})$\footnote{
It is discussed that the sensitivity might be improved to 
$O(10^{-(18 {\text -}19)})$ in the PRISM experiment \cite{Kuno:2013mha}.
}.

In our model, the branching ratio for the $\text{Au}$ target is predicted as \cite{Kitano:2002mt}
\begin{equation}
\begin{split}
   \text{BR}(\mu^- \text{Au} \to e^- \text{Au}) 
   &= 
   2.2 \times 10^{-13}  \left( \frac{g_X}{0.073}\right)^4  
   \left( \frac{100\,\text{TeV}}{m_{Z'}} \right)^4 
   \left( A^{l_L}_{12} \right)^2 
   \left| 1 + 0.58 A^{d_R^c}_{11} \right|^2,
\end{split}      
\end{equation}
which is close to the current upper bound at the SINDRUM-I\hspace{-1pt}I.  
The branching ratio for the Al target, which is a candidate target of 
COMET, Mu2e, and PRISM experiments, is evaluated as
\begin{equation}
\begin{split}
   \text{BR}(\mu^- \text{Al} \to e^- \text{Al}) 
   &= 
   6.3 \times 10^{-14}  \left( \frac{g_X}{0.073}\right)^4  
   \left( \frac{100\,\text{TeV}}{m_{Z'}} \right)^4 
   \left( A^{l_L}_{12} \right)^2 
   \left| 1 + 0.61 A^{d_R^c}_{11} \right|^2. 
\label{Eq:Al}   
\end{split}      
\end{equation}
The branching ratios for the other materials could be estimated as 
$O(10^{-13})$ as well, so that we expect that our model could be proved in 
the future experiments.

\subsection{Neutral Meson Mixing} \label{Sec:meson} 

The $Z'$ FCNCs contribute to the mass splitting and CP violation in neutral meson systems.
The UTfit collaboration analyzes the experimentally allowed ranges for the effective 
couplings of $4$-Fermi interactions~\cite{Bona:2007vi}.
We obtain the limits on the $Z'$ interaction as follows: 
\begin{eqnarray} \label{eq:CP}
   - 9.8 \times 10^{-3} < 
   \left( \frac{g_X}{0.073} \right)^2 
   \left( \frac{100\text{TeV}}{m_{Z'}} \right)^2 
   \text{Im}[(A^{d_R^c}_{12})^2]  
   &<& 1.6 \times 10^{-2},  \\
   \left( \frac{g_X}{0.073} \right)^2 
   \left( \frac{100\text{TeV}}{m_{Z'}} \right)^2 
   \left| \text{Re}[(A^{d_R^c}_{12})^2] \right|  
   &<& 3.4,   \\
   \left( \frac{g_X}{0.073} \right)^2 
   \left( \frac{100\text{TeV}}{m_{Z'}} \right)^2 
   \left| (A^{d_R^c}_{13})^2 \right|  
   &<& 81,   \\
   \left( \frac{g_X}{0.073} \right)^2 
   \left( \frac{100\text{TeV}}{m_{Z'}} \right)^2 
   \left| (A^{d_R^c}_{23})^2 \right|  
   &<& 3.9 \times 10^3.   
\label{Eq:bound_meson}         
\end{eqnarray}

The measurement of $K^0$-$\overline{K}^0$ oscillation is a strong
probe on both real and imaginary part of $(A^{d_R^c}_{12})^2$.
Especially, the CP violation gives a sever constraint on the FCNC as
we see in Eq.~(\ref{eq:CP}), so that the $Z'$ mass has to be heavier than
a few PeV, if $A^{d_R^c}_{12}$ possesses $O(1)$ CP phase.

\section{Conclusion and Discussion}
\label{sec4}

We have proposed an $SO(10)$ SUSY GUT, where the $SO(10)$ gauge
symmetry breaks down to $SU(3)_c \times SU(2)_L \times U(1)_Y\times
U(1)_{X}$ at the GUT scale and $U(1)_X$ is radiatively broken at the
SUSY-breaking scale.  In order to achieve the observed Higgs mass
around $126$~GeV and also to satisfy constraints on flavor- and/or
CP-violating processes, we assume that the SUSY-breaking scale is
$O(100)$~TeV, so that the $U(1)_X$ breaking scale is also
$O(100)$~TeV. In order to realize realistic Yukawa couplings, not only
${\bf 16}$-dimensional but also ${\bf 10}$-dimensional matter fields
are introduced.
The SM quarks and leptons are linear combinations of the ${\bf 16}$-
and ${\bf 10}$-dimensional fields so that the $U(1)_{X}$ gauge
interaction may be flavor violating.  We investigate the current
constraints on the flavor violating $Z'$ interaction from the flavor
physics and discuss prospects for future experiments.  Our model could
be tested in the flavor experiments, especially searches for the
$\mu$-$e$ conversion processes, even if the $Z'$ mass is $O(100)$~TeV.

In this paper, we did not mention the GUT mass hierarchy problem such
as the doublet-triplet splitting problem.  In fact, there is another
mass hierarchy between the singlet of ${\bf 16}_H$ and the other
components of ${\bf 16}_H$ in our model. The $Z'$ mass is given by the VEV
of the singlet, while other components reside around the GUT scale.  We
need more careful study on physics at the GUT scale to complete our
discussion.


\section*{Acknowledgments}
This work is supported by Grant-in-Aid for Scientific research
from the Ministry of Education, Science, Sports, and Culture (MEXT),
Japan, No. 24340047 (for J.H.), No. 23104011 (for J.H. and Y.O.), and No. 25003345 (for M.Y.). 
The work of J.H. is also supported by World Premier
International Research Center Initiative (WPI Initiative), MEXT,
Japan. 


\vspace{-1ex}

\begin{thebibliography}{99}
\bibitem{GG}
H.~Georgi and S.~L.~Glashow, 
Phys.\ Rev. \ Lett. {\bf 32}, 438 (1974).
\bibitem{su5}
  J.~R.~Ellis, S.~Kelley and D.~V.~Nanopoulos,
  Phys.\ Lett.\ B {\bf 249}, 441 (1990);
  J.~R.~Ellis, S.~Kelley and D.~V.~Nanopoulos,
  Phys.\ Lett.\ B {\bf 260}, 131 (1991).
\bibitem{so10}
  H.~Georgi,
  AIP Conf.\ Proc.\  {\bf 23}, 575 (1975);
  H.~Fritzsch and P.~Minkowski,
  Annals Phys.\  {\bf 93}, 193 (1975).

\bibitem{dim5protondecay}
  N.~Sakai and T.~Yanagida,
  Nucl.\ Phys.\ B {\bf 197}, 533 (1982).
  


\bibitem{FermionMass via HDO}
  H.~Georgi and C.~Jarlskog,
  Phys.\ Lett.\ B {\bf 86}, 297 (1979).

\bibitem{FermionMass via EH}
  J.~R.~Ellis and M.~K.~Gaillard,
  Phys.\ Lett.\ B {\bf 88}, 315 (1979).
  G.~Lazarides, Q.~Shafi and C.~Wetterich,
  Nucl.\ Phys.\ B {\bf 181}, 287 (1981).

\bibitem{Barr:1981wv}
  S.~M.~Barr,
  Phys.\ Rev.\ D {\bf 24}, 1895 (1981).

\bibitem{HSSUSY}
  N.~Arkani-Hamed and S.~Dimopoulos,
  JHEP {\bf 0506}, 073 (2005)
  [hep-th/0405159];
  G.~F.~Giudice and A.~Romanino,
  Nucl.\ Phys.\ B {\bf 699}, 65 (2004)
  [Erratum-ibid.\ B {\bf 706}, 65 (2005)]
  [hep-ph/0406088];
  N.~Arkani-Hamed, S.~Dimopoulos, G.~F.~Giudice and A.~Romanino,
  Nucl.\ Phys.\ B {\bf 709}, 3 (2005)
  [hep-ph/0409232];
  J.~D.~Wells,
  Phys.\ Rev.\ D {\bf 71}, 015013 (2005)
  [hep-ph/0411041];
  G.~F.~Giudice and A.~Strumia,
  Nucl.\ Phys.\ B {\bf 858}, 63 (2012)
  [arXiv:1108.6077 [hep-ph]];
    L.~J.~Hall and Y.~Nomura,
  JHEP {\bf 1201}, 082 (2012)
  [arXiv:1111.4519 [hep-ph]];
  M.~Ibe and T.~T.~Yanagida,
  Phys.\ Lett.\ B {\bf 709} (2012) 374
  [arXiv:1112.2462 [hep-ph]];
M.~Ibe, S.~Matsumoto and T.~T.~Yanagida,
  Phys.\ Rev.\ D {\bf 85} (2012) 095011
  [arXiv:1202.2253 [hep-ph]];
N.~Arkani-Hamed, A.~Gupta, D.~E.~Kaplan, N.~Weiner and T.~Zorawski,
  arXiv:1212.6971 [hep-ph].

\bibitem{GCU in split SUSY}
  J.~Hisano, T.~Kuwahara and N.~Nagata,
  Phys.\ Lett.\ B {\bf 723}, 324 (2013)
  [arXiv:1304.0343 [hep-ph]].

\bibitem{Hisano:2013exa}
  J.~Hisano, D.~Kobayashi, T.~Kuwahara and N.~Nagata,
  JHEP {\bf 1307} (2013) 038
  [arXiv:1304.3651 [hep-ph]].
\bibitem{InverseSeesaw}
R. N. Mohapatra and J. W. F. Valle, Phys. Rev. D{\bf 34} (1986) 1642; M. C. Gonzalez-Garcia and
J. W. F. Valle, Phys. Lett. B{\bf 216} (1989) 360; F. Deppisch and J. W. F. Valle, Phys. Rev. D{\bf 72} (2005)
036001 [hep-ph/0406040].

\bibitem{SO(10) breaking chain}
  J.~M.~Gipson and R.~E.~Marshak,
  Phys.\ Rev.\ D {\bf 31}, 1705 (1985);
  D.~Chang, R.~N.~Mohapatra, J.~Gipson, R.~E.~Marshak and M.~K.~Parida,
  Phys.\ Rev.\ D {\bf 31}, 1718 (1985);
  N.~G.~Deshpande, E.~Keith and P.~B.~Pal,
  Phys.\ Rev.\ D {\bf 46}, 2261 (1993).

\bibitem{PQ} 
  R.~D.~Peccei and H.~R.~Quinn,
  Phys.\ Rev.\ D {\bf 16}, 1791 (1977);
  R.~D.~Peccei and H.~R.~Quinn,
  Phys.\ Rev.\ Lett.\  {\bf 38}, 1440 (1977).

\bibitem{axion}
  J.~Preskill, M.~B.~Wise and F.~Wilczek,
  Phys.\ Lett.\ B {\bf 120}, 127 (1983);
  L.~F.~Abbott and P.~Sikivie,
  Phys.\ Lett.\ B {\bf 120}, 133 (1983);
  M.~Dine and W.~Fischler,
  Phys.\ Lett.\ B {\bf 120}, 137 (1983).


\bibitem{future}
  J.~Hisano, Y.~Muramatsu, Y.~Omura and M.~Yamanaka,
  in preparation. 


\bibitem{protondecay}
J.~Hisano, D.~Kobayashi and N.~Nagata,
  Phys.\ Lett.\ B {\bf 716} (2012) 406
  [arXiv:1204.6274 [hep-ph]];
J.~Hisano, D.~Kobayashi, Y.~Muramatsu and N.~Nagata,
  Phys.\ Lett.\ B {\bf 724} (2013) 283
  [arXiv:1302.2194 [hep-ph]].

\bibitem{Murakami:2001cs}
  B.~Murakami,
  Phys.\ Rev.\ D {\bf 65} (2002) 055003
  [hep-ph/0110095].




\bibitem{Bellgardt:1987du}
  U.~Bellgardt {\it et al.}  [SINDRUM Collaboration],
  Nucl.\ Phys.\ B {\bf 299} (1988) 1.




\bibitem{Blondel:2013ia}
  A.~Blondel, A.~Bravar, M.~Pohl, S.~Bachmann, N.~Berger, M.~Kiehn, A.~Schoning and 
  D.~Wiedner {\it et al.},
  arXiv:1301.6113 [physics.ins-det].





\bibitem{Bertl:2006up} 
  W.~H.~Bertl {\it et al.}  [SINDRUM II Collaboration],
  Eur.\ Phys.\ J.\ C {\bf 47}, 337 (2006).


\bibitem{Natori:2014yba} 
  H.~Natori [DeeMe Collaboration],
  Nucl.\ Phys.\ Proc.\ Suppl.\  {\bf 248-250}, 52 (2014).
  

\bibitem{Kuno:2013mha}
  Y.~Kuno [COMET Collaboration],
  PTEP {\bf 2013} (2013) 022C01.


\bibitem{Bartoszek:2014mya} 
  L.~Bartoszek {\it et al.}  [Mu2e Collaboration],
  arXiv:1501.05241 [physics.ins-det].




\bibitem{Kitano:2002mt}
  R.~Kitano, M.~Koike and Y.~Okada,
  Phys.\ Rev.\ D {\bf 66} (2002) 096002
   [Erratum-ibid.\ D {\bf 76} (2007) 059902]
  [hep-ph/0203110].
\bibitem{Bona:2007vi}
  M.~Bona {\it et al.}  [UTfit Collaboration],
  JHEP {\bf 0803} (2008) 049
  [arXiv:0707.0636 [hep-ph]].



\end{thebibliography}
\end{document}